\documentclass[aps,pra,twocolumn,floatfix]{revtex4-1}

\pdfoutput=1

\usepackage[utf8]{inputenc}
\usepackage{physics,amssymb,amsmath}
\usepackage{bm}
\usepackage{graphicx}
\usepackage[usenames,dvipsnames]{xcolor}
\usepackage[colorlinks,bookmarks=false,citecolor=blue,linkcolor=cyan,urlcolor=blue]{hyperref}
\usepackage{todonotes}
\graphicspath{{./figures/},{./prep/}}

\begin{document}

\title{Time-resolved observation of a dynamical phase transition of atoms in a cavity}
\author{T.~W.~Clark${}^{1*}$, A.~Dombi${}^{1*}$,  F.~I.~B.~Williams${}^1$,  \'A.~Kurk\'o${}^1$, J.~Fortágh${}^2$, D.~Nagy${}^1$, A.~Vukics${}^1$, P.~Domokos${}^1$}

\affiliation{${}^1$ Institute for Solid State Physics and Optics, Wigner Research Centre for Physics,  H-1525 Budapest P.O. Box 49, Hungary}

\affiliation{${}^2$ Physikalisches Institut, Eberhard Karls Universit\"at T\"ubingen, Auf der Morgenstelle 14, D-72076 T\"ubingen, Germany}

\begin{abstract}
The transparence of a laser-driven optical resonator containing an ensemble of cold atoms can have two distinct, robust states. Atoms in their initially prepared pure state blockade the transmission by detuning the cavity mode from the laser drive. The interacting system can, however, transition into an uncoupled state via a non-linear channel opening up in a critical run-away process toward a transparent bright phase.  The experiment enables a time-resolved observation of the dynamical transmission blockade breakdown phase transition as well as quantification of  enhanced fluctuations in the critical region.
\end{abstract}

\maketitle

Phase transitions are ubiquitous in macroscopic systems of interacting particles. The large size of macroscopic bodies generically inhibits investigation of the dynamics of a phase transition. Mesoscopic systems with controllable interaction between the particles open a route to study this fundamental phenomenon in a quantitative way. To this end, many-atom cavity QED systems represent an outstanding platform where the interaction between the components, \textit{i.e.} atoms and a few selected modes of the radiation field, is particularly well-controlled \cite{ritsch_cold_2013}. There are several cavity QED effects that can be cast into the class of phase transitions. One family is based on atomic self-organization \cite{domokos_collective_2002,black_observation_2003} when an ensemble of atoms in an optical resonator, illuminated by an external laser drive, can occupy various distinct spatial configurations depending on the intensity and frequency fine-tuning of the laser \cite{nagy_dicke-model_2010,baumann_dicke_2010,arnold_self-organization_2012,schmidt_dynamical_2014,klinder_dynamical_2015,leonard_supersolid_2017,kollar_supermode-density-wave-polariton_2017}. The other family of dynamical phase transitions is rooted in optical bistability \cite{lugiato_ii_1984,geng_universal_2020}, which occurs also for few atoms \cite{rempe_optical_1991,kerckhoff_remnants_2011,dombi_optical_2013}.  In the corresponding laser-driven cavity configuration, there are examples where bistability arises from the collective motion of an atomic cloud  \cite{gupta_cavity_2007} or Bose condensate  \cite{brennecke_cavity_2008}.

The ultimate quantum limit of phase transitions in the optical bistability configuration is the breakdown of the photon blockade  \cite{carmichael_breakdown_2015,dombi_bistability_2015,gutierrez-jauregui_dissipative_2018,pietikainen_photon_2019,rota_quantum_2019}. A cavity mode is driven resonantly by an external laser pump field which is transmitted through the empty cavity. However, when a single resonant atom is strongly coupled to the cavity mode, the system goes out of resonance with the pump, due to the vacuum Rabi splitting \cite{birnbaum_photon_2005}. The quantized energy eigenstates of the coupled system form a very unequally separated level structure \cite{schuster_nonlinear_2008}, such that the frequency mismatch of the driving laser with any of the transitions results in a blockade of the transmission. It has been shown that the blockade breaks down at very high intensities in the form of a phase transition \cite{vukics_finite_2019,brookes_critical_2021,curtis_critical_2021,reiter_cooperative_2020}.
The required regime of strong coupling constants is available only with circuit QED systems \cite{fink_observation_2017,mavrogordatos_strong-coupling_2019} using superconducting artificial atoms coupled to microwave resonators \cite{gu_microwave_2017}.  In optical resonators, and with a single atom, the photon blockade breakdown can only be realized far from the phase transition regime.

In this paper we present a many-atom variant of the photon-blockade breakdown. It takes place in an optical cavity with moderate electric dipole coupling to atoms. The many-atom enhancement  leads to a large collective coupling strength. The transmission of an originally resonant laser probe through the cavity can be suppressed by the frequency shift of the cavity mode due to a collective dispersive effect of the atoms. This is the mechanism for the transmission blockade.  As the atoms are sitting in the dark, the blockaded phase protects itself and is robust.  The alternative phase is that the atoms are all in a state decoupled from the cavity mode,  the cavity is filled with photons and resonant transmission of the drive laser can be observed. The change between them happens in a phase transition process which is continuously monitored by a photodetector at the cavity output. The transition is incited by fluctuations and driven by positive feedback. Our experiment enables time-resolved recording of the evolution of the order parameter during the transition. Moreover, we report on the observation of thermal photon fluctuations when the system is in between the two phases, and we demonstrate that the intensity of fluctuations diverges as a power law when the thermodynamic limit is approached.

\begin{figure*}
\includegraphics[width=1.8\columnwidth]{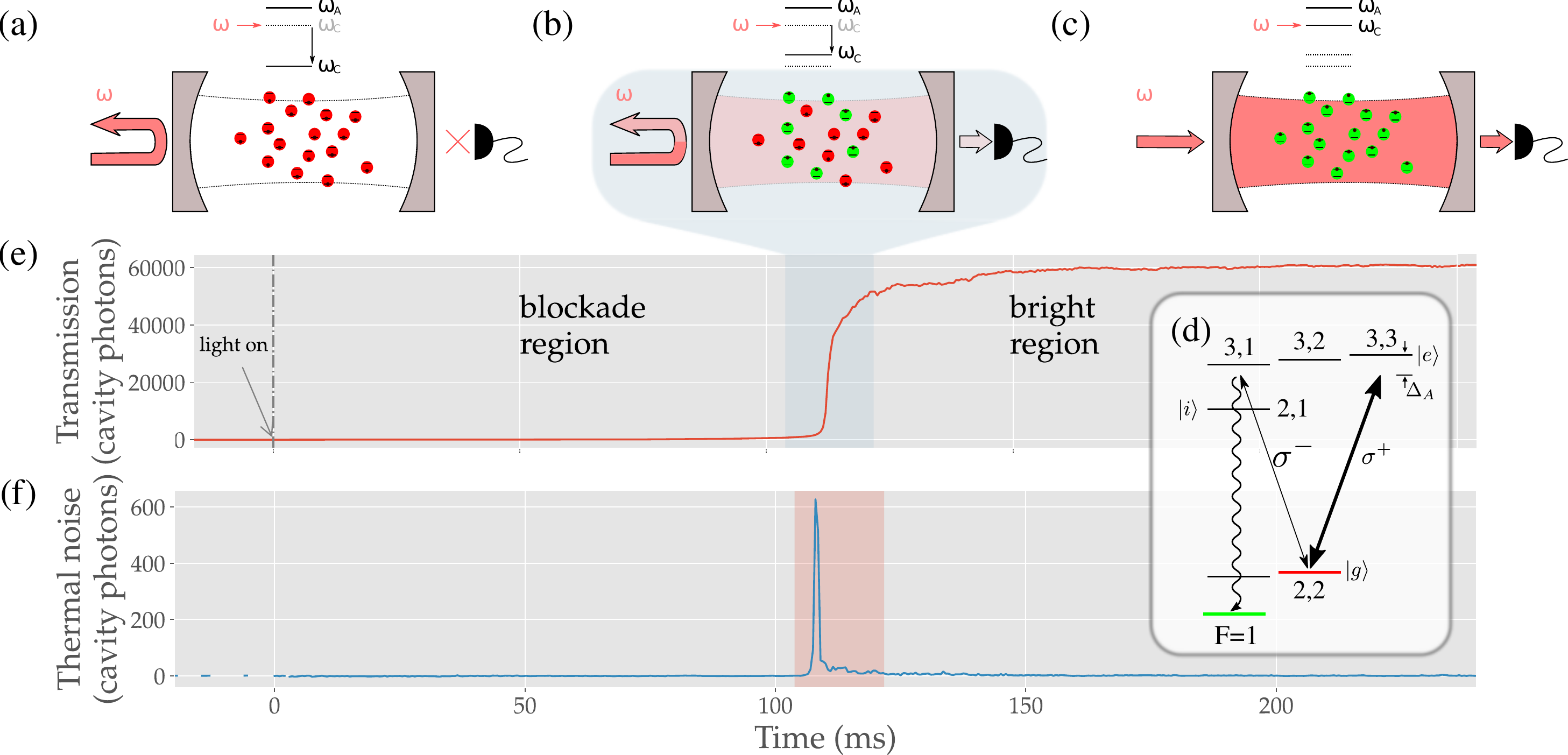}
\caption{Schematic representation of the transmission blockade breakdown phase transition. Atoms can be in  `red', (a), or `green', (c),  states, blocking or permitting the light transmission through the cavity, respectively. In the transition domain, (b), the atoms are in a mixture of red and green states. Upper level schemes show the cavity mode frequency with respect to the angular frequency of the pump laser, $\omega$ and in panel (d), red and green states are identified with the hyperfine states of ${}^{87}$Rb (only a part of the $5^{2}{\rm S}_{3/2} \leftrightarrow 5^{2}{\rm P}_{5/2}$ structure is shown). Far-off-resonance $\sigma^-$-polarized light provides an excitation path that assists the atoms' escape from the blockading state, $|g\rangle$, to the $F=1$ manifold of the electronic ground state. Atoms are first weakly excited to an intermediary state, $|i\rangle = (F, m_F) =( 2, 1)$, before spontaneously decaying to the manifold, which is optically dark with respect to the cavity mode. The time evolution of the transmitted intensity is plotted in (e), exhibiting the switch from blockaded to transparent phase around 100 ms after turning on the cavity drive. It is expressed in units of cavity photon number deduced from the detected photon flux. The transition is accompanied by the increase in cavity field fluctuations, represented in (f), in terms of thermal photon numbers extracted from the statistics of the transmitted light.}
\label{fig:TBB_summary} 
\end{figure*}
The system and the basics of the transmission blockade phase transition are schematically represented in Fig.~\ref{fig:TBB_summary}.
A single, standing-wave mode of a cavity with frequency, $\omega_C$, and linewidth, $\kappa$,  is externally driven by coherent laser light at a frequency, $\omega$. The transmission of the laser through the cavity exhibits a Lorentzian resonance which is modified if atoms are present in the cavity. Consider a number of atoms, ${\cal N}$, with electric dipole resonance, $\omega_A$,  which is far from the laser frequency, such that the atomic detuning,  $\Delta_A = \omega-\omega_A$, satisfies  $|\Delta_A|  \gg \gamma$, where  $\gamma$ is the  linewidth of the atomic resonance.  In this limit, the atoms act on the light field as a dispersive medium. Each atom in its electronic ground state, $|g\rangle$, shifts the frequency of the mode  by $\delta \cdot  | f(\mathbf{r}_j)|^2$ where $\delta= g^2/\Delta_A$, $g$ is the single photon Rabi frequency, $g=\sqrt{\tfrac{\omega_C}{2\epsilon_0\hbar {\cal V}}} d_{eg}$, and $d_{eg}$ is the atomic dipole moment. The second factor of the shift is the spatial mode function for atom $j = 1\ldots {\cal N}$. As the mode function, $f(\mathbf{r})$, is real and normalized to have a maximum of 1, the mode volume is ${\cal V} = \int {\rm d}^3 \mathbf{r} | f(\mathbf{r})|^2$. The frequency shift is additive and so the collective effect of the atoms gives a diminished transmission
\begin{equation}
\label{eq:transmission}
\frac{I_\text{out}}{I_0} = \frac{1}{\qty(\Delta_C-N  \delta)^2/\kappa^2 + 1} \; ,
\end{equation}
relative to the resonant transmitted intensity of the empty cavity, $I_0$. For resonant driving, $\Delta_C=\omega-\omega_C=0$, and a resonance shift much larger than the linewidth, $N \delta \gg \kappa$, the transmission is suppressed, which is the blockaded phase. The key variable governing the phase transition is the effective number of atoms, $N$, which depends both on the atomic positions and the internal state of the atoms,
\begin{equation}
\label{eq:N}
N = \sum_{j=0}^{\cal N} | f(\mathbf{r}_j)|^2 \cdot p_j \; .
\end{equation}
The internal state is represented by $p_j$, the difference in probability for the $j$th atom to occupy the ground or excited state respectively, $\text{Tr}\{\hat \rho \, (|g\rangle\langle g| - |e\rangle\langle e| ) \}$, where $|g\rangle \leftrightarrow |e\rangle$ labels the electric dipole transition coupled to the cavity mode. This concisely accounts for both a change in sign of the resonance, $-\delta$, due to population inversion and the actual number of atoms coupled to the mode, as optical pumping into dark states leads to $\text{Tr}\{\hat \rho \, (|g\rangle\langle g| + |e\rangle\langle e| ) \} \neq 1$.

Initially, all of the atoms are prepared in the state $|g\rangle$, such that $p_j=1$ for all $j$ (Fig.~\ref{fig:TBB_summary}a). After the probe light is turned on however, some light infiltrating into the cavity leads to a small atomic excitation into $|e\rangle$, and an even smaller component into another state, $|i\rangle$ (\textit{cf.} the level scheme in Fig.~\ref{fig:TBB_summary}d). From this latter state, the atoms can decay into a state decoupled from the cavity mode (`green' atoms  in Fig.~\ref{fig:TBB_summary}b). Both of these processes, in turn, reduce the variable $N$ and thus the  collective mode shift, letting more light enter the cavity. This positive feedback loop is closed, causing a system runaway into the fully transparent state (Fig.~\ref{fig:TBB_summary}c). The occurrence of the transmission blockade breakdown after a significantly long time (200 ms $\gg \gamma^{-1}, \kappa^{-1}$) and its associated dynamics are represented by the transmitted mean intensity in Fig.~\ref{fig:TBB_summary}e.

In our system, we used ${}^{87}$Rb atoms: first captured from vapour in an ultra-high vacuum chamber and then pre-cooled in a magneto-optical trap (MOT) above a high finesse optical resonator. The atoms were further cooled by polarization gradient cooling to reach typical temperatures of $T \sim 100 \mu$K. Following an optical pumping cycle, the magnetically polarized sample of cold ${}^{87}$Rb atoms in the $(F, m_F)=(2, 2)$ hyperfine ground state was loaded into a magnetic quadrupole trap. The magnetic trap center was shifted, in a controlled way, to transport the atoms vertically $\sim 1$ cm into the horizontally aligned cavity. The cavity is $l=15$ mm long and so has a relatively large access from the direction transverse to the propagation axis. The mode waist, $w=127 \mu$m, was an order of magnitude smaller than the size of the atomic cloud in this direction, placing approximately ${\cal N} \sim 10^5$ atoms within the cavity mode volume. The mode linewidth was measured to be $\kappa=  2\pi \cdot 3.22$ MHz (HWHM), and the single-atom coupling constant was calculated as $g= 2\pi \cdot 0.33$ MHz on the $(F, m_F)=(2, 2) \leftrightarrow (3,3)$ hyperfine transition of the D2 line.

Such conditions were achieved by driving the fundamental Gaussian mode of the resonator with an appropriate laser through the in-coupling mirror. The driving laser was locked to an atomic resonance and the resonator length was actively stabilized to the same atomic reference line via a transfer cavity at a far-detuned wavelength (805~nm). Thus the detuning, $\Delta_C$, was an actively controlled variable, set on resonance, $\Delta_C = 0$, and far below the F=2 $\leftrightarrow$ 3 atomic resonance by $\Delta_A=-2\pi \cdot 35$ MHz. The single-atom frequency shift was $\delta \approx 2\pi \cdot 3$ kHz, thus an effective number of atoms $N \approx 10^4$ led to a shift of the mode by more than $10 \kappa$ away from resonance. The transmission was blockaded under these conditions.

The magnetic quadrupole trap was centred in the cavity mode, \textit{i.e.} the mode was situated in the central plane of the trap where the magnetic field points radially outward from the symmetry axis. The atoms typically revolved around the (vertical) symmetry axis at a distance much larger than the mode waist. Within the cavity mode therefore, the atoms experienced a  magnetic field oriented parallel to the cavity axis. The quantization axis was thus aligned with the cavity axis, although pointing in opposite directions within each (longitudinal) half of the cavity mode. The circularly polarized light   injected into the cavity, $\sigma^+$,  excited the $(F, m_F)= (2, 2) \leftrightarrow (3, 3)$ closed-cycle transition with a Clebsch-Gordan coefficient equal to 1 in one half of the cavity. In the other half however, the light effectively had a $\sigma^-$ polarization with respect to the quantisation axis and weakly drove the $(F, m_F)= (2, 2) \leftrightarrow (3, 1)$ transition and, off-resonantly, the $(F, m_F)= (2, 2) \leftrightarrow (2, 1)$  transitions with Clebsch-Gordan coefficients of 1/15 and 1/3, respectively. This latter off-resonant excitation ($\Delta_A^\prime= 230$ MHz) by $\sigma^-$ light led to optical pumping into the $F=1$ manifold of the electronic ground state, which were dark states for the cavity field (\textit{cf.} Fig.~\ref{fig:TBB_summary}(d)). As this two-photon transition involved a virtual excitation of the state $|i\rangle$, intra-cavity intensity was needed. This constituted a non-linear decay channel for losing atoms from state $|g\rangle$, the state blockading the cavity transmission. Such an effect can underly the phase-transition-like switch from the ensemble of atoms in the  state $(F, m_F)= (2, 2)$ to the state $(F, m_F)= (1, m_F)$, with $m_F=0, 1$.

\begin{figure}
\includegraphics[width=0.81\columnwidth]{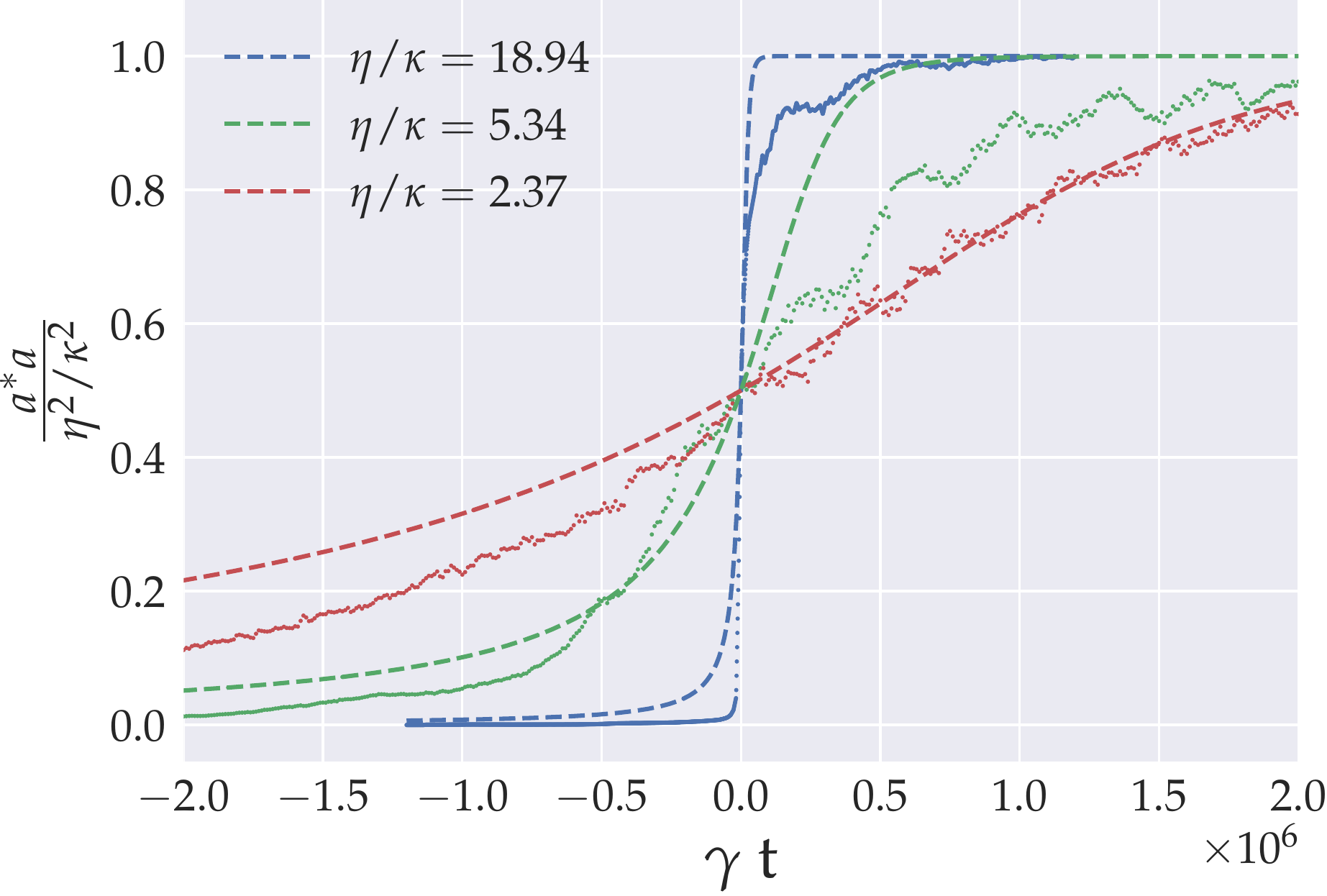}
\caption{The time evolution of the intracavity intensity around the phase transition for both  measurement data (solid line) and the mean-field simulation (dashed line). A selection of external laser drive powers, in units of the corresponding empty cavity photon number $(\eta/\kappa)^2$, are presented, where increasing drive power leads gradually from a crossover to the transmission blockade breakdown phase transition. After horizontally shifting the curves to have a common midpoint, the simplified model, with a single fitting parameter $\Gamma$, simultaneously accounts well for the slope of the transition  for all drive powers ($\Gamma=0.93 \cdot 10^{-3} \gamma$).}
\label{fig:intensities}
\end{figure}
A simple semiclassical model captures the phase transition dynamics. The usual atom-cavity interaction is complemented by an additional loss process with rate $\Gamma$  describing the escape to the dark states by spontaneous emission from the excited state. The mean-field approximation to the full quantum problem leads to the equations
\begin{equation}
\label{eq:meanfield}
\begin{split}
  \dot a &=  (i \Delta_C-\kappa ) a + gM + \eta,\\
 \dot M&= (i \Delta_A -\gamma -\Gamma ) M + g\left[N_e-N_g\right]a, \\
  \dot N_e&= - g\left[a^* M+ M^* a\right]-
            2 (\gamma + \Gamma)  N_e \textrm{ and }\\
  \dot N_g&=g\left[a^* M+M^* a \right]+
          2 \gamma N_e  \,,
\end{split}
\end{equation}
where $a$ is the complex amplitude of the cavity field mode driven by the effective amplitude, $\eta$. Concerning the other variables,  $M= {\cal N} \, \text{Tr}\left\{ |g\rangle\langle e | \right\}$ describes the atomic polarization and  $N_g =  {\cal N} \, \text{Tr}\left\{ |g\rangle\langle g | \right\}$, $N_e =  {\cal N} \, \text{Tr}\left\{ |e\rangle\langle e | \right\}$ the atomic populations. In this mean-field model the atoms are assumed to identically couple to the mode with an average coupling constant. The effective atom number in the transmission formula of Eq.~(\ref{eq:transmission}) is then $N=(N_g-N_e)/2$, where the factor $\tfrac12$ accounts for the reduction of the average coupling constant compared to its maximum. On integrating these equations from the appropriate initial conditions, \textit{i.e.} cavity vacuum, $a=0$, and all of the atoms in the ground state, $N_g= {\cal N}$, $N_e=M=0$, one can obtain the time evolution of  the transmitted intensity signal, $2\kappa |a|^2$, which serves as an order parameter for the phase transition.

This can be seen in Figure~\ref{fig:intensities}, where, focussing on the transition region, the slow cross-over from the blockaded transmission to the empty cavity phase ($N_g=N_e=0$) develops into ever faster switching on increasing the laser drive. Three different drive amplitudes are shown, spanning an intensity range of over two orders of magnitude. For the largest power (blue lines), the mean field solution is matched to the experimental data by using the escape rate, $\Gamma$, as the only fitting parameter and the number of atoms set to $N=10^4$.  For the same value of the escape rate, $\Gamma = 0.93 \cdot 10^{-3}\gamma$, the slope of the transition around the midpoint exhibits good agreement between measurement and simulation simultaneously for the other two drive powers. Suggesting that the essence of the phase transition dynamics is well captured by Eqs.~(\ref{eq:meanfield}).

\begin{figure}[tb]
\includegraphics[width=\columnwidth]{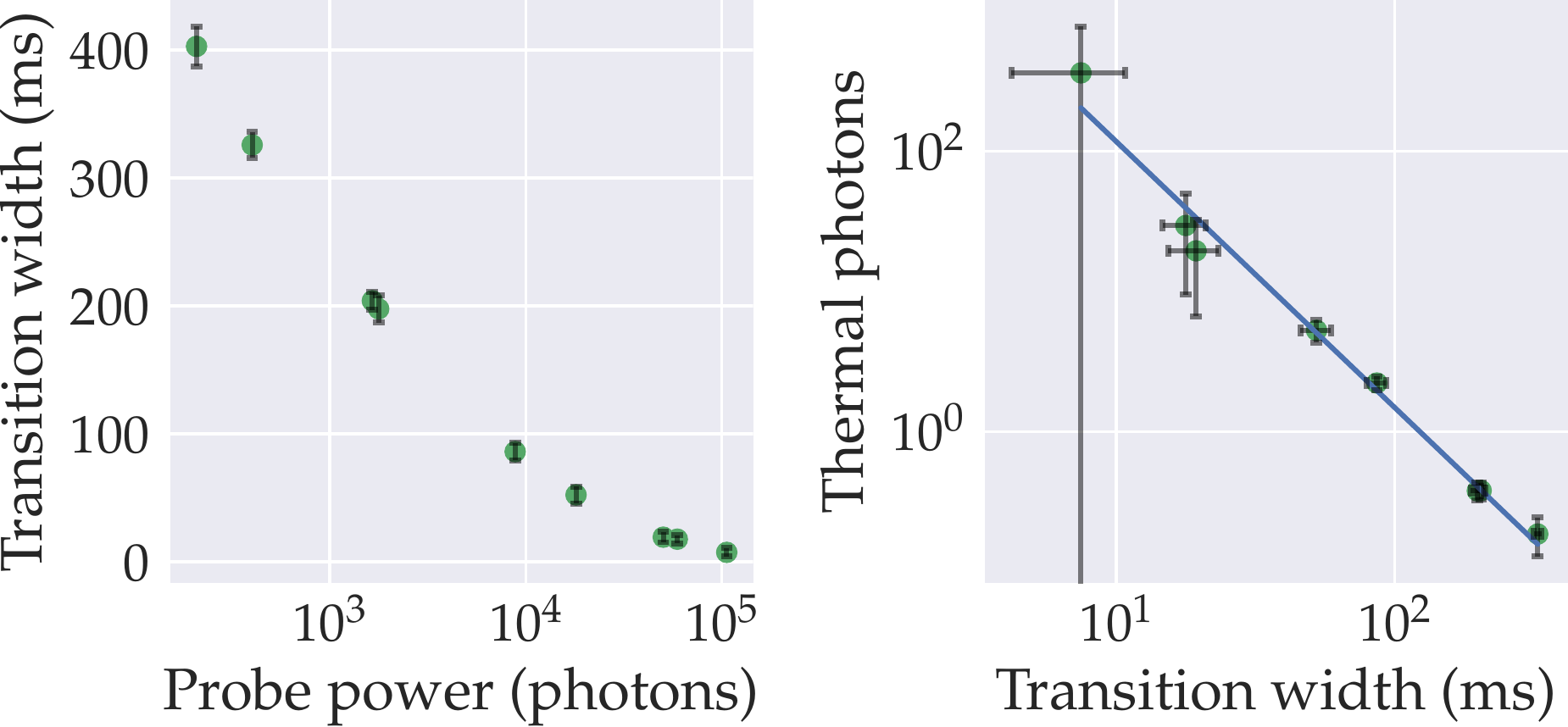}
\caption{Left: The width of the transition as a function of the laser drive power, highlighting a finite-size feature of the transition in the transmission blockade breakdown. Right: The scaling of thermal photon number as a function of transition width, where the latter indicates distance from the thermodynamic limit. The fitted power law suggests an exponent of $-1.9 \pm 0.1$.}
\label{fig:finitesize}
\end{figure}
With increasing laser drive power the transition happens more quickly, as plotted in Fig.~\ref{fig:finitesize}. Here, the transition width was defined as the time taken for the transmitted intensity to rise from 10\% to 90\% of the resonant empty cavity transmission (\textit{cf.} the shaded region of the sample trajectory in Fig.~\ref{fig:TBB_summary}(e)). In order to approach the thermodynamic limit, the enhanced drive power should be accompanied by increasing the number of atoms so that the collective dispersive effect counteracts the larger incoming light intensity. On doing this scaling, the transition tends to an instantaneous change. In our experiment the atom number is not varied, however, the transition width can be operationally used as a measure for how far the system is from the thermodynamic limit.

Our experiment reveals a generic feature of phase transitions beyond the mean-field level, \textit{i.e.} the emergence of enhanced fluctuations in the course of the transition \cite{goes_quantum_2020}. The intensity of cavity field fluctuations was extracted from the running variance of the recorded transmission signal with 500 $\mu$s time resolution \footnote{see Supplementary Material}. The variance can be connected to the $g^{(2)}$ intensity correlation function of the single-mode field \cite{carmichael_open,foster_intensity_1998,overbeck_silicon_1998} which expresses the enhancement of the cavity field fluctuations with respect to the Poissonian statistics.  As can be seen in Fig.~\ref{fig:TBB_summary}(f), the cavity photon fluctuations exhibit a sharp peak in the time evolution, just at the moment where the order parameter transitions from the blockaded phase. This excess noise can be expressed in terms of a thermal photon number by using the ansatz for the state of the cavity mode that it is a statistical mixture of coherent states with a Gaussian distribution, $P_{\rm th, disp} (\alpha) = \frac{1}{\pi n_\text{th}} \exp\left(-{|\alpha - \beta|^2}/{n_{\rm th}}\right) $. This is the $P$-function of a displaced thermal state, with mean field denoted by the complex amplitude $\beta$ and where the distribution width, $n_{\rm th}$, can be interpreted as the number of thermally distributed photons. For this mixed state, the  intensity correlation function obeys $g^{(2)}(0)  = 2 - \frac{|\beta|^4}{( n_{\rm th} + |\beta|^2)^2}$. This value lies between 1 and 2 for a coherent ($n_{\rm th}=0$) and thermal ($\beta=0$) state, respectively. In the course of the transition, the mean-field amplitude evolves from $\beta=0$ to $\beta=\eta/\kappa$, as shown in Fig.~\ref{fig:TBB_summary}(e). The width of the distribution, $n_{\rm th}$, also changes during the transition, and its time resolved evolution was derived from the measured data, as exemplified in Fig.~\ref{fig:TBB_summary}(f).

The thermal noise is related to the internal dynamics of the atoms and its description is beyond the scope of our mean-field model (\ref{eq:meanfield}). In the blockaded regime the transmitted field must be close to a vacuum state. In the transparent phase the transmitted field statistics is expected to retain the Poissonian statistics of the laser source. In between, when the atoms are partially excited, the atomic state can be a statistical mixture of states $|g\rangle$ and $|e\rangle$, which is encoded via the distribution of the probabilities, $p_j$, in the effective atom number, $N$, in Eq.~(\ref{eq:N}).  This mixture amounts to additional statistical features in the detected field above the Poissonian noise.

Finally, we show that the fluctuations generated in the transition increase as a power law as the thermodynamic limit is approached.  In Figure~{\ref{fig:finitesize}}, the integrated thermal photon number, $n_{\rm th}$, is plotted as a function of the transition width in a log-log scale together with a power law fit. This function represents a finite-size scaling. Theoretical confirmation of the measured exponent, $-1.9 \pm 0.1$, requires more involved modelling. Nevertheless, the good agreement with the fit over two orders of magnitude confirms a characteristic feature of phase transitions, \textit{i.e.} the power law divergence of the fluctuations as the thermodynamic limit is approached.   We can conclude that the experimentally observed breakdown of the transmission blockade corresponds  to a finite-size realization of a genuine phase transition.

\section*{Acknowledgements}

T.~W.~Clark and A.~Dombi contributed equally to this work. This work was supported by the National Research, Development and Innovation Office of Hungary (NKFIH) within the Quantum Technology National Excellence Program  (Project No. 2017-1.2.1-NKP-2017-00001) and the Quantum Information National Laboratory of Hungary. D. Nagy and A. Vukics were supported  by the J\'anos Bolyai Fellowship of the Hungarian Academy of Sciences.

\bibliography{TBB1}

\end{document}